\documentstyle[aps,prl,multicol,epsf]{revtex}
\begin{document}
\def\CC{{\rm\kern.24em \vrule width.04em height1.46ex depth-.07ex
\kern-.30em C}}
\def\P{{\rm I\kern-.25em P}}
\def\RR{{\rm
         \vrule width.04em height1.58ex depth-.0ex
         \kern-.04em R}}

\draft
\title{ Computation on a Noiseless Quantum Code and Symmetrization}
\author{Paolo Zanardi$^{1,2}$
}
\address{
$^{1}$ Istituto Nazionale per la Fisica della Materia (INFM) \\
$^{2}$ Institute for Scientific Interchange  Foundation, \\Villa Gualino,
Viale Settimio Severo 65, I-10133 Torino, Italy\\
}
\date{\today}
\maketitle
\begin{abstract}
Let ${\cal H}$ be the state-space of a quantum computer
coupled with the environment by a set of error 
operators spanning a Lie algebra ${\cal L}.$
Suppose ${\cal L}$ admits a noiseless quantum
code i.e., a subspace ${\cal C}\subset{\cal H}$ annihilated by ${\cal L}.$
We show that a universal set of gates
over $\cal C$ is obtained by  any generic
pair of ${\cal L}$-invariant gates.
Such gates - if not available from the outset - can be obtained
by resorting to a symmetrization with respect to the group generated by ${\cal L}.$ 
Any computation can then be performed completely within the coding
decoherence-free subspace.
\end{abstract}
\maketitle
\begin{multicols}{2}

Universality in  the quantum computation  framework \cite{QC}
means that any unitary operator  can be approximated
to any desired extent by a sequence of elementary transformations in a set $G$.
Such elementary transformations, referred to as the {\em universal quantum gates},
are the ones  that  are supposed to be available to the experimenter. 
It is physically plausible
that if one is able to implement the unitary $U_i$ then $U_i^{-1}=U_i^\dagger$
should be implementable as well. Moreover it is also natural
to assume $\openone\in G.$ In this case the set of realizable operations (sequences of elements of $G$)
form a subgroup of the group $U({\cal H})$ of unitary transformations over the quantum state-space $\cal H.$
Such subgroup has to be {\em dense} in $U({\cal H})$ \cite{dense}.

In reference \cite{SETH} it is shown that a universal set of gates over  $\cal H$
can be  realized by  a pair of unitaries $\exp(i\,t_l\, H_l)\,(l=1,2)$
simply by requiring that the Lie algebra generated by the  $i\,H_l$'s
amounts to  the whole  Lie algebra $u({\cal H})$ of anti-hermitian operators.
Even though this result is just existential, it  is quite remarkable in that it tells us
that if one is able to switch on a {\em generic}   pair of Hamiltonians $H_1$ and $H_2,$ 
in principle any unitary transformation can be obtained.
Notice that here one is {\em not} assuming  the system to be multi-partite
i.e., a tensor product structure on $\cal H.$
With this assumption  constructive results can be obtained \cite{UG}.

In this note we address the problem of universality 
within a  {\em Error Avoiding Code.} 
The latter basically is a decoherence-free subspace in which
quantum information can be reliably stored  \cite{CHI}, \cite{ZARA}, \cite{LID}.
In view of general results on universal quantum computation,
this issue might seem rather trivial  \cite{CHI},  \cite{LID}.
Indeed if  one has at disposal an universal set of gates for $\cal H$
then even all the transformations over the code $\cal C$ can be realized,
for example, in the following way.
Suppose $\cal C$ encodes $N$ (logical) qubits, then ${\cal C}=\varphi(\CC^{2^N})$
where $\varphi$ is a unitary map.
In the $N$-partite case a  universal set of gates $\{U_i\}_{i\in I}$
is known (single qubit operations and
almost any two-qubit operation \cite{UG}) and  its image $\{\hat\varphi( U_i)\}_{i\in I},\,
(\hat\varphi( U_i) :=\varphi\,U_i\,\varphi^\dagger)$
is universal on $\cal C$ \cite{uni}.
Unfortunately this approach is only formal: in real systems the $\hat \varphi( U_i)$'s [more precisely
their unitary extension to the whole $\cal H$] have to be realized by means of the
physical  gates one can switch on.
The point is that, {\em in general, the  necessary sequences of gates
will draw  the state out of $\cal C$ during the intermediate steps.}
In this way the system is exposed to dissipation and decoherence
i.e., computational errors.

We shall  characterize the generic universal set of gates
over the code by analysing the algebraic structures associated with $\cal C$
at the operatorial level.
Moreover we will discuss how to ideally realize
such gates   by means of  the recently introduced 
symmetrization procedures \cite{SYMM},\cite{VIO}.
In this way one can, in principle, obtain  a controlled dynamics completely within
the code.

{\em Error Avoiding}--
We  begin by  recalling   some very basic notions about algebras and representations \cite{CORN}
relevant to the theory of error avoiding codes \cite{ZARA}.
[We  denote
by $\mbox{End}({\cal H})$ the (associative and Lie) algebra of linear operators
on $\cal H.$
When ${\cal H}\cong {\CC}^d$
this space is isomorphic with the set $M(d, {\CC})$ of complex $d\times d$ matrices.]
Let $\cal L$ be a (semisimple, finite-dimensional) Lie algebra and
$\rho\colon {\cal L}\rightarrow \mbox{End}({\cal H})$ a representation
of $\cal L$ in the Hilbert space $\cal H$ ($\mbox{dim}\,{\cal H}=d < \infty$).
This map extends naturally to  a representation 
(also denoted by $\rho$)
of  the associative algebra with unit $\cal A$ generated by $\cal L$ \cite{UEA}.
Let us consider the decomposition of $\rho$ according to the irreducible representations $\rho_J$
of $\cal L$ (from now on $\cal L$-{\em irreps})
$\rho=\oplus_J n_J \rho_J,$ where $n_J\in{\bf{N}}$ denotes the multiplicity
of the $J$-th $\cal L$-irrep $\rho_J.$ Of course  
$\sum_J n_J\,d_J=d,\,\mbox{dim}\,\rho_J= d_J.$
The state-space splits accordingly
\begin{equation}
{\cal H}=   \bigoplus_J {\cal H}_J\cong 
\bigoplus_J {\CC}^{n_J}\otimes {\CC}^{d_J}.
\label{split}
\end{equation}
The latter isomorphism is  due to the fact that the  subspace ${\cal H}_J,$
of vectors transforming according to $\rho_J,$ 
 is made of $n_J$ copies of a $d_J$-dimensional  irrep subspace.
There  follows that
$
{\cal L}_\rho:= \rho({\cal L})
\subset \bigoplus_J {\openone}_{n_J}\otimes M(d_J, {\CC}). 
$
Since ${\cal A}$ acts irreducibly in each factor ${\CC}^{d_j},$
its image under $\rho$  coincides with the whole algebra of operators, and this
provides the further isomorphism
\begin{equation}
{\cal A}_\rho:= \rho({\cal A})\cong \bigoplus_J  
\openone_{n_J}\otimes M(d_J, {\CC}).
\label{alg-split}
\end{equation}
This equation shows that the  space of operators commuting with  ${\cal A}_\rho$
i.e, the {\em commutant}  
is given by
\begin{equation}
{\cal A}_\rho^\prime\cong \bigoplus_J M(n_J, {\CC})\otimes\openone_{d_J}.
\label{central}
\end{equation}
Note that ${\cal A}^\prime_\rho$ coincides with the centralizer $Z({\cal L}_\rho)$ 
 of operators commuting with ${\cal L}_\rho.$
We restate in the present setting the results of ref.
\cite{ZARA}.  

{\em Definition}
Let $J=0$ be the label for the $1$-dimensional $\cal L$-irrep, then
${\cal C}:= {\cal H}_0\cong {\CC}^{n_0}\otimes  {\CC}\cong
{\CC}^{n_0},$  the  singlet sector of $\cal L$
 will be referred to as the {\em code} \cite{ZARA}.

Since ${\cal L}$ is semisimple the $1$-d irreps are trivial. There follows that
$
{\cal C}=
\{|\psi\rangle\in{\cal H}\colon \rho(x)|\psi\rangle=0,\forall x\in{\cal L}\}.
$
The error avoiding theorem stems  from this observation

{\em{Proposition 1}} 
Let the  dynamics over ${\cal H}\otimes{\cal H}_E$
(the second factor represents the environment) be
generated by the Hamiltonian $H=H_S\otimes\openone + \openone\otimes H_E +H_I,$
where 
$H_I=\sum_{\alpha} S_{\alpha}\otimes B_{\alpha}.$
If $[H_0, {\cal L}_\rho]\subset {\cal L}_\rho$   and $\{ S_{\alpha}\}\subset {\cal A}_\rho$
then the induced subdynamics over $\cal C$ is unitary.
If also $H_0\in {\cal A}_\rho,$
then any observable  of ${\cal A}_\rho^\prime$ is a constant of the motion.

In physical applications one typically starts from the {\em error generator}
operators $S_\alpha,$ then ${\cal L}$ (${\cal A}$) is defined as the minimal Lie (associative) subalgebra
of $\mbox{End}({\cal H})$ containing them.
[The algebras  are concretely given, the representation index $\rho$
can be dropped.]
Isomorphisms (\ref{alg-split}) and (\ref{central}) are recovered by
the central decomposition of $\cal A.$
The assumption that the system-environment coupling
satisfies the  conditions stated in Prop. 1 for an algebra representation
such that $n_0(\rho)\neq 0,$ is a strong symmetry constraint.
It selects a special class of correlated decoherence 
interactions. {\em Generically the algebra spanned by the error generators
acts irreducibly over} $\cal H$.
In other terms in the generic case the ``super-selection'' structure
described by eq. (\ref{alg-split}) collapses to a single term
coinciding with the whole operator algebra i.e., ${\cal A}\cong\mbox{End}({\cal H}).$

To exemplify the situation
we now consider the case    of collective decoherence of a quantum register
made of $N$ $d$-dimensional cells.
The relevant algebra is   ${\cal L}= {\em sl}(d, {\CC})$
and $\rho$ is the $N$-fold tensor power
of the defining representation (i.e., ${\cal H}\cong ({\CC}^d)^{\otimes\,N}$).
Then  ${\cal A}^\prime_\rho$ is the image under the natural representation
$\nu$ of the {\em group algebra} \cite{gr-al} ${\CC}{\cal S}_N$ of the symmetric group ${\cal S}_N$
over a $N$-partite state-space
[i.e., $\nu(\sigma)\otimes_{j=1}^N|j\rangle=\otimes_{j=1}^N|\sigma(j)\rangle,\,\sigma\in{\cal S}_N.$]
For the qubit case ($d=2$) the multiplicities are given by the elementary
angular momentum theory \cite{CORN}:
$
n_N(J)=(2\,J+1)\,N!/[(N/2+J+1)!\,(N/2-J)!], 
$
and the dimension of the irreps is $d_J=2\,J+1\,(J\in{\bf{N}}/2)$ \cite{cat}
Notice that one has ${\cal A}^{\prime\prime}_\rho=\nu(\CC{\cal S}_N)^\prime={\cal A}_\rho.$
Indeed trivially ${\cal A}_\rho\subset {\cal A}^{\prime\prime}_\rho,$ moreover
$\mbox{dim}\, {\cal A}_\rho=\sum_J d_J^2=1/6(N+3)\,(N+2)\,(N+1),$
is the dimension of the subspace of the totally symmetric operators. 

Now we  characterize the  operators
over the code.

{\em Proposition 2} 
i) ${\cal A}_\rho|_{\cal C}\cong {\CC}.$
[The operators of ${\cal A}_\rho$ restricted to the code
are proportional to the identity.]

ii) ${\cal A}^\prime_\rho|_{{\cal H}_J}\cong
M(n_J, {\CC})\otimes {\openone}_{d_J}$; 
in particular the elements of the commutant  restricted to the code 
span the algebra of operators over $\cal C.$ i.e.,
\begin{equation}
{\cal A}^\prime_\rho|_{{\cal C}}= \mbox{End}({\cal C})
\cong M(n_0, {\CC}).\label{restri}
\end{equation}
{\em Proof.}
i)  Stems from 
 the fact that any element of ${\cal A}_\rho$ is given by 
a term proportional to the identity plus a polynomial
in the $\rho(x_j),$  the $x_j$'s being a basis for $\cal L.$

ii) Is a consequence of eq. (\ref{central}).
$\hfill\Box$

Before stating  the main result of this paper let us briefly discuss
the case in which the  operators $S_\alpha$ in the interaction Hamiltonian
are hermitian and span an {\em abelian algebra} i.e., $[S_\alpha,\,S_{\alpha^\prime}]=0.$
Altough this (non semi-simple)  situation is pretty singular, it can be physically realized
in purely decohering systems \cite{ZA}.
The state-space splits according to the common eigenspaces of the $S_\alpha$'s,
${\cal H}=\oplus_{{{s}}=1}^M {\cal C}_{ {{s}} }.$ 
[$|\psi\rangle\in{\cal C}_s\Rightarrow S_\alpha\,|\psi\rangle=\lambda_{s,\alpha}\,|\psi\rangle$.]
If $\Pi_{{{s}}}$ denotes the projector over ${\cal C}_{ {{s}}},$
one has $\Pi_{{{s}}}\,\Pi_{{{s}}^\prime}=
\delta_{ {{s}}, {{s}}^\prime}\Pi_{{{s}}},\,
\sum_{{{s}}} \Pi_{{{s}}}=\openone.$
Then the relevant group structure
is given now by ${\cal Z}_2^M$ acting on $\cal H$
via the direct product representation 
$\rho\colon {\bf{m}}\mapsto
\exp\left(i\,\pi\,{\bf{m}}\cdot{\bf{P}} \right )= \openone -2\,{\bf{m}}
\cdot{\bf{P}}. 
$
where $ {\bf{m}}:=(m_1,\ldots,m_M)\in{\cal Z}_2^M,\,{\bf{P}}:=
(\Pi_1,\ldots,\Pi_M).$

 The characterization result of prop.  2 clearly suggests that a solution to the problem
of computation within the code  {\em can be obtained by taking a set of gates contained
in ${\cal A}_\rho^\prime.$}
If it happens that one has at disposal from the beginning
$\cal L$-invariant operators, then 
any transformation over the code can be implemented without leaving
it at any step.

{{\em Theorem 1}}
Suppose that one can switch  on two hamiltonians $H_i\in{\cal A}_\rho^\prime \,(i=1,2),$ 
then in the generic case the set of gates $\{e^{i\,H_i}\}$ is universal over $\cal C.$

{\em Proof.} It follows  from point ii) of Prop. 2
by means of  the same arguments used in ref. \cite{SETH} for proving
generic universality results.
$\hfill\Box$

Since ${\cal A}_\rho^\prime$ is a proper subspace of $\mbox{End}({\cal H}),$
generically a gate is not ${\cal L}_\rho$-invariant.
For instance single qubit operations, apart from the identity, transform
according the adjoint representation of $sl(2,\CC)$ and therefore
do not leave the code invariant in collective decoherence.
One has  to face the problem of how to build  ${\cal L}_\rho$-invariant
gates from non invariant ones.
This issue will be  addressed hereafter.

{\em Symmetrizing}--
We present  first a formal procedure
for getting  unitary operators over the code, that shows a new potential applications of the ideas
discussed 
in ref. \cite{SYMM}.
For the sake of concreteness 
we consider   the case of collective $sl(d,\,\CC)$-decoherence,
in which one has $\mbox{End}({\cal C})=\nu(\CC{\cal S}_N)|_{\cal C}.$
The state-space is enlarged to ${\cal H}\otimes \CC{\cal S}_N$
and  
\begin{equation}
|0\rangle:=\frac{1}{\sqrt{N!}}\sum_{\sigma\in{\cal S}_N} |\sigma\rangle,\;
W_\nu:=\sum_{\sigma\in{\cal S}_N} \nu_\sigma\otimes |\sigma\rangle\langle \sigma|.
\end{equation}
Suppose we want to implement the unitary  operator
$X=\sum_\sigma x_\sigma \nu_\sigma\in \CC{\cal S}_N$
over a code element $|\psi\rangle,$ then we can:

I) Prepare the initial state
$|\Psi_0\rangle:= |\psi\rangle\otimes |X\rangle.$
Where $|X\rangle:= \sum_\sigma x_\sigma |\sigma\rangle$
is normalized in view of the  unitarity of $X.$

II) Apply $W_\nu:$ 
$W_\nu\,|\Psi_0\rangle=\sum_{\sigma\in{\cal S}_N} x_\sigma \,\nu_\sigma|\psi\rangle
\otimes |\sigma\rangle.
$

III) Measure $\openone\otimes |0\rangle\langle 0|$ and discard the ancilla.
With probability $1/N!$ we obtain
$ X|\psi\rangle.$

This algorithm  requires extra space resources (in the present case an exponentially large
ancillary space), the capability of performing the highly non trivial
operation $W_\rho$ and finally it has a probability of success smaller than one.
All this can make it hardly implementable. 
 
We  discuss now 
a different scheme for generating the transformation over the code
that does not require extra space resources.
The key idea is that the required $\cal L$-invariant universal gates 
can be obtained from a universal set of gates in $\cal H$
by the symmetrization procedure for quantum evolutions introduced in \cite{SYMM}.

Let ${\cal G}_\rho$ be the Lie group
generated  by ${\cal L}_\rho.$
The operators commuting with this group, or ${\cal G}_\rho$-invariant, 
are the same that commute with ${\cal L}_\rho$ i.e., 
${\cal A}_\rho^\prime =Z({\cal L}_\rho)=Z({\cal G}_\rho)$ \cite{ele}.
Such elementary remark implies  that in the {\em continuos} case symmetrizing
with respect to the algebra $\cal L_\rho$
is the same as  symmetrizing  respect to its associated group ${\cal G}_\rho.$
These last issues have been recently addressed in 
\cite{SYMM}, and  we recall some of the notions involved.

The  ${\cal G}_\rho$-symmetrizing map $\pi_{\rho}$  is given by 
\begin{equation}
\pi_{\rho} \colon 
X\rightarrow \int_{\cal G} dy\, \rho(y)\,X\,\rho(y)^\dagger, 
\label{pi}
\end{equation}
where $X\in\mbox{End}({\cal H})$ and $dy$ is an invariant (Haar) measure over $\cal G.$
The basic properties of $\pi_{\rho}$
are now established  

{\em Proposition 3}
i) $\pi_{\rho}$ is a projector;
ii)
$
\mbox{Im}\, \pi_{\rho}:=
\pi_{\rho} (\mbox{End}({\cal H}))=Z({\cal G}_\rho);
$
iii) $ \pi_{\rho}({\cal L}_\rho)= 0.$

{\em Proof.}
i) One has to verify that $\pi_{\rho}^2=\pi_{\rho}$
and $\pi_{\rho}^\dagger=\pi_{\rho}.$
Both the properties follow by straightforward calculations
using the invariance of the Haar measure.

ii) The property  $\pi_{\rho}(X)\in Z({\cal G}_\rho),\,\forall X\in \mbox{End}({\cal H})$
i.e., $\mbox{Im}\,\pi_{\rho}\subset  Z({\cal G}_\rho),$ is also verified
by an explicit calculation.
The opposite inclusion is immediately proved in that, from eq. (\ref{pi}),
$X\in Z({\cal G}_\rho)\Rightarrow \pi_{\rho}(X)=X.$

iii) The operators of $\pi_{\rho}({\cal L}_\rho)\subset {\cal A}_\rho^\prime$
 have in particular to commute
with all the elements of ${\cal L}_\rho.$
Since $\pi_{\rho}$ is a linear combination of inner automorphisms of ${\cal L}_\rho,$
one has  $\pi_{\rho}({\cal L}_\rho)\subset {\cal L}_\rho$ as well.
From the semisimplicity of ${\cal L}_\rho$ it follows  that its
 centre is trivial.

When $\cal A$ is  abelian  the map  (\ref{pi}) takes a particularly simple form:
$\pi_{\rho}(X)=
\sum_{s=1}^M \Pi_s\,X\,\Pi_s.$

From the practical point of view it is  extremely important that the continuos average in eq. 
(\ref{pi}) can be replaced by an averaging  over a
 finite-order group ${\cal E}=\{e_j\}_{j=1}^{|{\cal E}|}$ of unitary transformation 
i.e., 
$\pi_{\rho}(X)=\pi_{\cal E}(X):=|{\cal E}|^{-1}\sum_{j=1}^{|{\cal E}|} e_j\,X\,e_j^\dagger.$
Such   ${\cal E}$  exists for any finite-dimensional operator algebra; 
it can be  obtained by (unitarily) gluing together the finite order symmetrizing groups ${\cal E}_J$
that can be found  
for each   irrep subspace ${\cal H}_J.$
This fact can be seen as follows.

Let 
 $\{e_j^{J}\}_{j=1}^{d^2_J}$ be a   basis of unitary transformations of 
the operator algebra over the second
 factor of each irrep space ${\cal H}_J\cong \CC^{n_J}\otimes\CC^{d_J}.$
For any $d_J$ such basis  can be chosen in such a way that it generates 
a {\em finite} subgroup ${\cal E}^{J}$ of $U(d_J)$  that acts irreducibly over $\CC^{d_J}$ \cite{KNI}.
Any operator $X$ commuting 
with ${\cal E}^{J}$ commutes in particular with the $e_j^J$'s,
since the latter span $M(d_J,\,\CC),$ there follows that $X$ is 
 proportional to the identity $\openone_{d_J}.$
One can unitarily 
extend  the action  of $g_K\in{\cal E}^K$ to the whole direct sum  (\ref{split})
by
$\tilde g_K(\sum_J \phi_J\otimes \psi_J) := \phi_K\otimes g_K\,\psi_K +\sum_{J\neq K} \phi_J\otimes \psi_J.$
Let $\{|J, n, l\rangle\}$ be the  basis of $\cal H$ defining the last isomorphisms of eq. 
(\ref{split}) i.e., $J$ labels the irreps appearing in the decomposition of $\rho,$
$n=1,\ldots,n_J,\;l=1,\ldots,d_J.$  
The following one-to-one map is a representation 
of the product group $\prod_J \tilde {\cal E}^J$ in ${\cal H},$ 
$ g:=\prod_J \tilde g_J\mapsto \hat g$ where
$
\langle J, n, l|\,\hat g\,|J^\prime, n^\prime, l^\prime\rangle=
\delta_{J,J^\prime}\,\delta_{n,n^\prime}\, (g_J)_{l,l^\prime}.
$
By an explicit evaluation of $\pi_{\cal E}$ it can be checked that a  finite  symmetrizing group for ${\cal A}_\rho$ is  given by 
${\cal E}=\{\hat g\}.$ 
This result can be understood by observing that:
 a) triviality within each ${\CC}^{d_J}$-factor
is obtained in view of the definition of the ${\cal E}_J$'s; b)
terms coupling different irreps $J$ and $J^\prime$ are killed, by the group averaging  in that
$(\sum_{g\in{\cal E}^J} g)\,{\CC}^{d_J}=0$ i.e., no invariant vectors exist.

To exemplify this general construction let us consider the case of $N=2$ qubits
collectively decohering.
The relevant algebra ${\cal A}_\rho\cong sl(2\,\CC)$  is generated by the global spin operators
$S^\alpha=\sigma^\alpha\otimes\openone+\openone\otimes\sigma^\alpha,\,(\alpha=\pm,z)$
and ${\cal H} \cong \CC^4\cong \CC^3\oplus \CC$
(triplet plus singlet sector).
One finds ${\cal E}=\{ Q^n\,P^m\}_{n,m=0}^2$ where $Q:=\exp(i\,2\pi/3\,S^z)$
and $P:=\exp[2\,\pi/\sqrt{3}\,( {2}^{-1/2}\,S^+-2^{-1}\,S^{+\,2}-\mbox{h.c.})].$
The  next lemma follows from  proposition 3 and eq. (\ref{restri}) 

{\em {Proposition 4}}
For any unitary operator $X$ over $\cal C$ there exists an Hamiltonian over $\cal H$
 such that 
$X=\exp[ i\,\pi_{\rho}(H)|_{\cal C}].$

The operator $H$ is not unique; there is a sort of gauge freedom. 
For example, from point iii) of Prop. 3 it follows that
one can add to $H$ any element of ${\cal L}_\rho.$ 
More generally any  $H^\prime$ such that $\pi_{\rho}(H)=\pi_{\rho}(H^\prime)$
will be good as well. Obviously this latter relation defines an equivalence relation 
$\sim$ among pairs of operators such that  
$
\mbox{End}({\cal H})/\sim= \mbox{End}({\cal H})/\mbox{Ker}\,\pi_{\rho}
\cong {\cal A}_\rho^\prime.
$
Each equivalence class (or {\em orbit})
 is specified by $d^2-\sum_J n_J^2=\mbox{dim}\,\mbox{Ker}\,\pi_{\rho}$ parameters.
We are now in the position to prove 

{\em {Theorem} 2}
Given two generic gates $U_i=e^{i\,t_i\,H_i}\in U({\cal H})\,(i=1,2)$
then their $\cal G_\rho$-symmetrization is an universal set of gates
over the code $\cal C.$

{\em Proof.}
 Two Hamiltonians $H_i\in i\,u({\cal H})$ generically fulfill the  condition
of reference \cite{SETH} for universal quantum computation over $\cal H.$
In the (generic) case in which  they do not belong to the same orbit 
i.e., $\pi_{\rho}(H_1)\neq
\pi_{\rho}(H_2)$ one has that  $\{\pi_{\rho}(H_i)|_{\cal C} \}_{i=1,2}$
generates the whole $u({\cal C}).$
Therefore any $X\in U({\cal C})$ can be approximated arbitrarily well
by sequences of $X_i=e^{i\,t_i\,\pi_{\rho}(H_i)|_{\cal C}};$
but according to reference \cite{SYMM} such operators can be realized
by a symmetrization procedure for  $U_i=e^{i\,t_i\,H_i}.$ 
$\hfill\Box$

In other words: {\em any evolution in the singlet sector $\cal C$ can be obtained by a restriction
to $\cal C$ of
the ${\cal G}_\rho$-symmetrization of an evolution over $\cal H.$ }
The key point is that, being now the infinitesimal generators ${\cal G}_\rho$-invariant,
the computation is in the ideal case performed  {\em completely}
within the decoherence-free subspace.
Notice that the gauge freedom discussed above can be helpful from the point of view    
of this scheme; indeed one might choose in the orbits the optimal hamiltonians
with respect to the criterion of physical realizability.
The  evolution generated by $\tilde H:=\pi_{\rho}(H)$ can be viewed as a sort  of continuos projection
over $\cal C$ of a state evolving according $H.$
This is very easily understood by considering the action of $\tilde H$
over a state  $|\psi\rangle\in{\cal C}$:
${\tilde H}\,|\psi\rangle =\int_{\cal G} dg \rho_g\,H\, \rho_g^\dagger\,|\psi\rangle
= \int_{\cal G} dg \rho_g\,H\,|\psi\rangle=\tilde \pi_\rho\,H\,|\psi\rangle.
$
Here we have used the fact that $\cal C$ is nothing but the identity $\cal G$-irrep space
and introduced the projector $\tilde \pi_\rho:= \int_{\cal G} dg \rho_g$ over it \cite{SYMM}.
This sort of quantum Zeno effect \cite{ZEN} is achieved, without performing any
measurement, by means of a { purely unitary control of the system dynamics} during the gate
operation. 

Let $\tau_{op}$ be the (typical) time-scale over which the unitary transformation corresponding
to a group elements of  ${\cal E}$
can be realized, and 
 $\tau_p$  the time interval between the pulses associated with  different
group elements.  
Here we hace been  tacitly assuming that 
$\tau_{op}\ll\tau_{p}\le |{\cal E}|\,\tau_p\ll
\tau_{c}=\omega_c^{-1},$ where $\omega_c$ denotes the cut-off frequency
of the system-environment dynamics \cite{VIO}.
Otherwise also the application of the symmetrizing ``pulses'' $\rho_g$'s,
that might  pull the state out of the code, would introduce
themselves errors. 
Let us stress that,
since the the errors generators $S_\alpha$ belong by definition to ${\cal A}_\rho,$
from point i) of Prop. 2, it follows that the symetrization procedure during
 gating results in a decoherence suppression \cite{SYMM},\cite{VIO}.
Finally notice that, as far as the gate is symmetrized, there is no constraint on the global
gating time.

{\em Summary}--
In this paper we have shown how to  achieve universal  computation
completely  constrained to a quantum error avoiding code $\cal C.$
The code is associated to a (Lie) algebra ${\cal L}_\rho$ of error generators, 
coupling the computational subsystem with its environment.
A generic pair of  unitary transformations commuting with ${\cal L}_\rho$ 
are sufficient to obtain arbitrary computations on $\cal C.$
If such  ${\cal L}_\rho$-invariant gates are not available from the outset
they can be ideally obtained  by symmetrizing a generic 
(non ${\cal L}_\rho$-invariant)  pair of gates
with respect to the group generated  by ${\cal L}_\rho.$
The symmetrization procedure requires that one is able to
switch on   interactions, corresponding to a finite subgroup, with a frequency
greater than the inverse of the fastest time-scale of system-environment dynamics.  
This result allows in principle for a complete fault-tolerant
quantum computation with error avoiding codes.

 I thank M. Rasetti for useful discussions and critical reading of the manuscript.
Elsag-Bailey for financial support.

\end{multicols}

\begin{references}
\bibitem{dense}
The set  of operators
over $\cal H$ is  endowed with the Hilbert space topology
generated by the  Hilbert-Schimdt scalar product  
$<A,\,B>:=\mbox{tr} (A\,B^\dagger), \,\|A\|:=\sqrt{<A,A>}.$
\bibitem{uni} Let $\cal T$ (${\cal T}_\varphi$) the group generated by the $U_i$'s
 ($\hat\varphi(U_i)$'s).
 $\hat\varphi$ is an  algebra isomorphism then $\hat\varphi (\cal T)= {\cal T}_\varphi.$
Moreover, from continuity, $\overline{\cal T}_\varphi=\hat\varphi(\overline{\cal T})=\hat\varphi
(U(\CC^{2^N}))=U({\cal C}).$
\bibitem{UEA} Let $\{x_j\}$ a basis of $\cal L,$ then $\cal A$ is defined as the
linear space of (formal) polynomials in the $x_j$'s. The latter satisfy the relations
$x_j\,x_i-x_i\,x_j=\sum_k c_{ji}^k\, x_k,$ where $\{c_{ij}^k\}$ are the structure constants
of $\cal L,$ and $x_j^0:=\openone.$
\bibitem{gr-al} Given a group $\cal G$ of finite order,  its group algebra $\CC{\cal G}$
is the   vector space generated by  complex combinations of elements of $\cal G.$
Multiplication is  introduced by linear extension of the group
operation.
\bibitem{ele}
Indeed if $X\in Z({\cal L}_\rho)$ and $g=\exp( y)\,(y\in{\cal L}_\rho)$
one has $g\,X=g\,X\,g^\dagger\,g=\mbox{Ad}g( X)\,g=\exp( \mbox{ad} y)(X)\,g= X\,g$
in that $[X,\,y]=0\,(\forall y\in{\cal L}_\rho).$
Conversely suppose $X\in Z({\cal G}_\rho)$ and $y\in{\cal L}_\rho.$
The elements of the $1$-parameter subgroup generated by $y,$  $g_t:=\exp( i\,t\,y)\,(t\in\RR),$
satisfy $[g_t,\,X]=0\, \forall t;$ then $\partial_t [g_t,\,X]|_{t=0}\sim [y,\,X]=0.$
\end{references}
\end{document}